\documentclass[aps,notitlepage,twocolumn,showpacs,showkeys,preprintnumbers,amsmath,amssymb]{revtex4-1}

\usepackage{graphicx}
\usepackage{dcolumn}
\usepackage{bm}

\renewcommand{\rho}{\varrho}
\renewcommand{\phi}{\varphi}
\renewcommand{\vec}{\mathbf}

\begin{document}

\title{Addendum to ``Free-energy functional of the Debye-H\"{u}ckel model of simple fluids''}

\author{R. Piron$^1$ and T. Blenski$^2$}

\affiliation{$^1$CEA, DAM, DIF, F-91297 Arpajon, France.}

\affiliation{$^2$Laboratoire ``Interactions, Dynamiques et Lasers'', UMR 9222, CEA-CNRS-Universit\'e Paris-Saclay, Centre d'\'Etudes de Saclay, F-91191 Gif-sur-Yvette Cedex, France.}

\date{\today}

\begin{abstract}
In previous publications \cite{Piron16,Blenski17}, the authors have proposed Debye-H\"{u}ckel-approximate free-energy functionals of the pair distribution functions for one-component fluid and two-component plasmas. These functionals yield the corresponding Debye-H\"{u}ckel integral equations when they are minimized with respect to the pair distribution functions, lead to correct thermodynamic relations and fulfill the virial theorem. In the present addendum, we update our results by providing simpler functionals that have the same properties. We relate these functionals to the approaches of Lado~\cite{Lado73} and of Olivares and McQuarrie~\cite{Olivares76}. We also discuss briefly the non-uniqueness issue that is raised by these results.
\end{abstract}


\maketitle

\section{Introduction}
In his paper \cite{Lado73}, Lado performs the calculation of a correction to the free-energy of a reference system, starting from the Debye-Kirkwood charging relation \cite{Kirkwood35}. If one disregards any reference system, the result can be viewed as an hypernetted chain (HNC) excess-free-energy functional of the pair distribution function. The same expression had also been previously derived by Morita and Hiroike \cite{Morita60}. Since then, this expression was often used in its functional interpretation (see, for instance \cite{Lado83,Ayadim09}).

In Ref.~\cite{Piron16}, our derivation of a Debye-H\"{u}ckel (DH) excess-free-energy functional was based on the same starting point as Ref.~\cite{Lado73}. It appeared to us later that, following even more closely the derivation of Ref.~\cite{Lado73}, one can obtain a different free-energy functional. The latter has the same thermodynamical properties while having a simpler expression, which is also more similar to the expression of Ref.~\cite{Lado73} for the HNC free-energy. This expression can also be recovered from the method described by Olivares and McQuarrie in \cite{Olivares76}. However, their method is focused on the construction of generating functionals and does not in itself provide systematically a free-energy functional.

In this addendum, starting from a brief summary of Lado's derivation, we derive this alternative DH free-energy functional in the one-component and multi-component cases. Each time, we show that it leads to the same thermodynamics as our previously-proposed functionals. We also discuss how the methods of \cite{Lado73} and \cite{Olivares76} complement each other. Finally, we try to shed some light on the non-unicity issue which is raised by the present work.

\section{Key steps of Lado's calculation}
 The derivation of the HNC free-energy given in \cite{Lado73}, as well as ours in \cite{Piron16}, starts from the charging relation:
\begin{align}
\frac{A_\text{eq}^\xi(\rho,T)}{V}=\frac{\rho^2}{2}\int_0^\xi d\xi' \int d^3r
\left\{h_\text{eq}^{\xi'}(r)u(r)\right\}
\label{eq_charging}
\end{align}
where, $A_\text{eq}^\xi/V$ is the equilibrium free-energy \footnote{For the sake of consistency with our previous publications \cite{Piron16, Blenski17}, and unlike Lado \cite{Lado73}, we will work with the renormalized excess free energy per unit volume.} of the simple fluid having inter-particle interaction potential $\xi u(r)$, and $h_\text{eq}^\xi(r)+1$ is the equilibrium pair distribution function ($h_\text{eq}^\xi(r)$ is called the equilibrium correlation function). This relation is fulfilled for the exact equilibrium quantities. In the present context, we require it to hold for the approximate equilibrium quantities, stemming from an approximate theory such as HNC or DH. The approximate equilibrium pair distribution function is such that the following closure relation holds:
\begin{align}
\ln(h_\text{eq}^\xi(r)+1)=&-\beta\xi u(r)+h_\text{eq}^\xi(r)\nonumber\\
&-c\left\{h_\text{eq}^\xi(r');r\right\}-b_\text{approx}\left\{h_\text{eq}^\xi(r');r\right\}
\label{eq_equilibrium}
\end{align}
with $b_\text{approx}\left\{h^\xi(r');r\right\}$ denoting the approximate bridge function corresponding to the chosen model. $c\left\{h(r');r\right\}$ is the direct correlation function, which we regard as a functional of $h(r)$ defined through the Ornstein-Zernike (OZ) relation:
\begin{align}
c\left\{h(r');r\right\}=h(r)+\rho\int d^3r'\left\{h(r')c\left\{h(r'');|\vec{r}-\vec{r}'|\right\}\right\}
\label{eq_ornstein_zernike}
\end{align}

Performing the derivative with respect to $\xi$ of the equilibrium relation \eqref{eq_equilibrium}, we obtain easily:
\begin{align}
&h_\text{eq}^\xi(r)\beta u(r)=\nonumber\\
&\frac{\partial}{\partial\xi}
\left(\frac{(h_\text{eq}^\xi(r))^2}{2}-(h_\text{eq}^\xi(r)+1)c\left\{h_\text{eq}^\xi(r');r\right\}
-\beta\xi u(r)
\right)\nonumber\\
&+c\left\{h_\text{eq}^\xi(r');r\right\}\frac{\partial h_\text{eq}^\xi(r)}{\partial\xi}
-(h_\text{eq}^\xi(r)+1)\frac{\partial}{\partial\xi}b\left\{h_\text{eq}^\xi(r');r\right\}
\end{align}
Using again Eq.~\eqref{eq_equilibrium} in the first term of the latter equation right-hand-side, we get:
\begin{align}
&h_\text{eq}^\xi(r)\beta u(r)=\nonumber\\
&\frac{\partial}{\partial\xi}
\left(\frac{(h_\text{eq}^\xi(r))^2}{2}+h_\text{eq}^\xi(r)\beta\xi u(r)-(h_\text{eq}^\xi(r)+1)
\right.\nonumber\\&\left.
\times(h_\text{eq}^\xi(r)-\ln(h_\text{eq}^\xi(r)+1)-b\left\{h_\text{eq}^\xi(r');r\right\})
\vphantom{\frac{1}{2}}\right)\nonumber\\
&+c\left\{h_\text{eq}^\xi(r');r\right\}\frac{\partial h_\text{eq}^\xi(r)}{\partial\xi}
-(h_\text{eq}^\xi(r)+1)\frac{\partial}{\partial\xi}b\left\{h_\text{eq}^\xi(r');r\right\}
\label{eq_lado_method}
\end{align}
Using this equation in the charging relation Eq.~\eqref{eq_charging}, the first term is straightforwardly integrated over the charging parameter $\xi$. In order to integrate the second term, we switch to the Fourier space and use the OZ relation (i.e. the definition of $c\left\{h^\xi(r');r\right\}$), in order to show that:
\begin{align}
&\int d^3r \left\{ c\left\{h^\xi(r');r\right\}\frac{\partial h^\xi(r)}{\partial\xi} \right\}\nonumber\\
&=\frac{\partial}{\partial\xi}\frac{1}{\rho^2}\int \frac{d^3k}{(2\pi)^3}
\left\{ \rho h^\xi_k-\ln(1+\rho h^\xi_k) \right\}
\end{align}
Here, we used the following definition of the Fourier transform of a function $f(r)$:
\begin{align}
\int d^3r \left\{ f(r) e^{i\vec{k}.\vec{r}} \right\}
\end{align}

In the case of the HNC model, which is considered in Lado's paper, the bridge function is disregarded: $b\left\{h(r');r\right\}=0$. Taking $\xi=1$, this leads to the approximate free-energy:
\begin{widetext}
\begin{align}
\frac{A_\text{eq}^\text{HNC}(\rho,T)}{V}=&\frac{\rho^2}{2\beta}\int d^3r \left\{ \left(\frac{(h_\text{eq}(r))^2}{2}+h_\text{eq}(r)\beta u(r)-(h_\text{eq}(r)+1)(h_\text{eq}(r)-\ln(h_\text{eq}(r)+1))\right) \right\}
\nonumber\\
&+\frac{1}{2\beta}\int \frac{d^3k}{(2\pi)^3}
\left\{ \rho h_{\text{eq},k}-\ln(1+\rho h_{\text{eq},k}) \right\}
\label{eq_hnc_functional}
\end{align}
\end{widetext}
This expression can be viewed as the HNC-equilibrium value of a functional of $h(r)$: $A_\text{eq}^\text{HNC}(\rho,T)=A^\text{HNC}\left\{h(r);\rho,T\right\}|_\text{eq}$. That is: $A^\text{HNC}\left\{h(r);\rho,T\right\}$ is defined from Eq.~\eqref{eq_hnc_functional} right-hand-side, treating $h_\text{eq}(r)$ as a variable. Moreover, it can be checked \emph{a posteriori} that minimization of $A^\text{HNC}$ with respect to $h(r)$ yields the HNC closure (i.e. Eq.~\eqref{eq_equilibrium} with $b\left\{h(r');r\right\}=0$).

Taking the problem from the other side, one can first search for the functionals of $h(r)$ that yields the OZ equation with a chosen closure relation, here: the HNC closure. Such a systematic approach is exposed in \cite{Olivares76}. Once these are found, one can then pick among them a functional that fulfills the charging relation. We will further comment on this point in Sec.~\ref{sec_Olivares}.

\section{Alternative Debye-H\"{u}ckel free-energy functional\label{sec_DH_functional}}
The DH integral equation for a simple fluid is:
\begin{align}
h_\text{eq}^\xi(r)=-\beta\xi u(r)-\rho\beta\xi\int d^3r'\left\{h_\text{eq}^\xi(r')u(|\vec{r}-\vec{r}'|)\right\}
\end{align}
This is equivalent to the OZ relation Eq.~\eqref{eq_ornstein_zernike} with the closure:
\begin{align}
c\left\{h_\text{eq}^\xi(r');r\right\}=-\beta\xi u(r)
\label{eq_DH_closure}
\end{align}
From Eq.~\eqref{eq_equilibrium}, one can thus write the corresponding bridge function as:
\begin{align}
b\left\{h_\text{eq}^\xi(r');r\right\}=h_\text{eq}^\xi(r')-\ln(h_\text{eq}^\xi(r)+1)
\end{align}
It turns out that the last term in Eq.~\eqref{eq_lado_method} can then be readily rewritten as:
\begin{align}
(h_\text{eq}^\xi(r)+1)\frac{\partial}{\partial\xi}b\left\{h_\text{eq}^\xi(r');r\right\}
=\frac{1}{2}\frac{\partial (h_\text{eq}^\xi(r))^2}{\partial\xi}
\end{align}
We end up with:
\begin{align}
\frac{A_\text{eq}^\text{DH}(\rho,T)}{V}
=&\frac{\rho^2}{2\beta}\int d^3r \left\{h_\text{eq}(r)\beta u(r) \right\}
\nonumber\\
&+\frac{1}{2\beta}\int \frac{d^3k}{(2\pi)^3}
\left\{ \rho h_{\text{eq},k}-\ln(1+\rho h_{\text{eq},k}) \right\}
\label{eq_DH_functional}
\\
=&\left.\frac{A^\text{DH}\left\{h(r);\rho,T\right\}}{V}\right|_\text{eq}
\end{align}
Like in previous section, $A^\text{DH}\left\{h(r);\rho,T\right\}$ is defined from Eq.~\eqref{eq_DH_functional} right-hand-side, treating $h_\text{eq}(r)$ as a variable.
Here again, one can readily check that minimization of $A^\text{DH}$ with respect to $h(r)$ yields the DH closure Eq.~\eqref{eq_DH_closure}. 

In Ref.~\cite{Olivares76}, the mean spherical model, which is closely related to the DH model, is addressed in order to find a generating functional of the direct correlation function. However, one can use the results of \cite{Olivares76} in order to obtain the functional of Eq.~\eqref{eq_DH_functional}. This is discussed in the next section.

The most direct way of checking the thermodynamical relations obtained using Eq.~\eqref{eq_DH_functional} is to compare the equilibrium value of $A^\text{DH}$ with that of our previous expression, given in Eq.~(19) of \cite{Piron16}. Let us first recall this expression:
\begin{align}
&\frac{A^{\text{DH}\,\prime}\left\{h(r);\rho,T\right\}}{V}
\nonumber\\
&=
\frac{\rho}{\beta}
\int \frac{d^3k}{(2\pi)^3}\left\{
\left(1+\frac{1}{\rho\beta u_k}\right)
\left(1-\frac{\ln\left(1+\rho \beta u_k\right)}{\rho\beta u_k}\right)
\right.\nonumber\\&\hphantom{=}\times\left.
h_k\left(\frac{h_k}{2}+\beta u_k+\frac{\rho\beta}{2}h_k u_k\right)
\right\}
\label{eq_DH_functional_2016}
\end{align}
Inserting the DH equilibrium value of $h_k$:
\begin{align}
h_{\text{eq},k}=\frac{-\beta u_k}{1+\rho\beta u_k}
\end{align}
we get after a few simplifications:
\begin{align}
&\frac{A^{\text{DH}\,\prime}\left\{h_\text{eq}(r);\rho,T\right\}}{V}\nonumber\\
&=\frac{1}{2\beta}\int \frac{d^3k}{(2\pi)^3}\left\{-\rho\beta u_k+\ln\left(1+\rho\beta u_k\right)\right\}
\label{eq_DH_eq_free_energy_expr}
\end{align}
On the other hand, it can be easily checked that, at equilibrium, Eq.~\eqref{eq_DH_functional} reduces to the same expression as Eq.~\eqref{eq_DH_eq_free_energy_expr}.

The identity, in the sense of thermodynamical functions, of both free-energy functionals, when taken at equilibrium, immediately yields the identity of all thermodynamical functions, i.e. derivatives of the equilibrium free energy with respect to the thermodynamic parameters $\rho$, $T$, and also with respect to $u(r)$\footnote{In the latter case, one views any thermodynamic function as a functional of $u(r)$. However, for the sake of conciseness, we did not write explicitly the $u(r)$ dependency of the thermodynamic functions. Any time it is useful we mention in the text whether a function depends or not on $u(r)$.}.

Moreover it is worth noting that the expression for the internal energy can be obtained more easily from Eq.~\eqref{eq_DH_functional} than from Eq.~\eqref{eq_DH_functional_2016}.
\begin{align}
\frac{U_\text{eq}^\text{DH}(\rho,T)}{V}=\left.\frac{\partial}{\partial\beta}\left(\frac{\beta A^\text{DH}\left\{h(r);\rho,T\right\}}{V}\right)\right|_\text{eq}
\end{align}
\begin{align}
\frac{\partial}{\partial\beta}\left(\frac{\beta A^\text{DH}\left\{h(r);\rho,T\right\}}{V}\right)
=\frac{\rho^2}{2}\int d^3r\left\{h(r) u(r)\right\}
\end{align}

It is also easier to show that the virial theorem is fulfilled:
\begin{align}
P_\text{eq}^\text{DH}(\rho,T)=\left.\rho^2\frac{\partial}{\partial\rho}\left(\frac{A^\text{DH}\left\{h(r);\rho,T\right\}}{\rho V}\right)\right|_\text{eq}
\end{align}
\begin{align}
\rho^2\frac{\partial}{\partial\rho}&\left(\frac{A^\text{DH}\left\{h(r);\rho,T\right\}}{\rho V}\right)
=\frac{\rho^2}{2}\int \frac{d^3k}{(2\pi)^3}\left\{h_k u_k\right\}
\nonumber\\
&+\frac{1}{2\beta}\int \frac{d^3k}{(2\pi)^3}\left\{\ln\left(1+\rho h_k\right)-c_k\left\{h(r)\right\}\right\}
\end{align}
\begin{align}
P_\text{eq}^\text{DH}=&\frac{U_\text{eq}^\text{DH}}{V}
+\frac{1}{2\beta}\int \frac{d^3k}{(2\pi)^3}\left\{
\rho\beta u_k-\ln\left(1+\rho\beta u_k\right)\right\}\\
=&P_\text{virial}^\text{DH}
\end{align}
where the expression for $P_\text{virial}^\text{DH}$ can be checked in \cite{Piron16}, Eq.~(16).

As is discussed in \cite{Piron16}, the thermodynamical consistency of the obtained expression is deeply related to the fact that we required Eq.~\eqref{eq_charging} to hold at equilibrium.

Although the functional of Eq.~\eqref{eq_DH_functional} leads to the same thermodynamics as that of Eq.~\eqref{eq_DH_functional_2016}, they still differ as functionals. Among the differences between the two is that the functional of Eq.~\eqref{eq_DH_functional} has the following derivative:
\begin{align}
\frac{\delta}{\delta u(r)}\frac{A^\text{DH}}{V}=\frac{\rho^2}{2}h(r)
\label{eq_derivative_u}
\end{align}
whatever $h(r)$ is considered, whereas this relation is only fulfilled at equilibrium with the functional of Eq.~\eqref{eq_DH_functional_2016}.
As is noted in \cite{Fantoni03}, appendix B, having Eq.~\eqref{eq_derivative_u} to be fulfilled at equilibrium is essential to the thermodynamic consistency of the free-energy functional. This equation indeed follows from the charging relation. However, fulfilling Eq.~\eqref{eq_derivative_u} for all $h(r)$, is a sufficient, but not a necessary condition for $A^\text{DH}$ to be a free-energy functional.

\section{Relation to the Olivares-McQuarrie approach\label{sec_Olivares}}
In Ref.~\cite{Olivares76}, Olivares and McQuarrie propose a somehow systematic method in order to find generating functionals for the integral-equation models based on the OZ relation.
The final purpose of this approach is thus to find functionals $\mathcal{F}\left\{h(r)\right\}$ such that the integral equation of the model is equivalent to:
\begin{align}
\frac{\delta \mathcal{F}\left\{h(r)\right\}}{\delta h(r)}=0
\label{eq_mini_F}
\end{align}
They choose to first search for a functional $\mathcal{C}\left\{h(r)\right\}$ such that the OZ relation is equivalent to:
\begin{align}
\frac{\delta \mathcal{C}\left\{h(r)\right\}}{\delta h(r)}=-\rho^2c(r)
\end{align}
that is:
\begin{align}
\frac{\delta \mathcal{C}\left\{h(r)\right\}}{\delta h_k}=\frac{-\rho^2 h_k}{1+\rho h_k}
\end{align}
This leads to:
\begin{align}
\mathcal{C}\left\{h(r)\right\}
=\mathcal{C}^*+\int \frac{d^3k}{(2\pi)^3}\left\{\rho h_k - \ln\left(1+\rho h_k\right)\right\}
\end{align}
$\mathcal{C}^*$ being a constant with respect to $h(r)$. Then, a functional $\mathcal{F}\left\{h(r)\right\}$ can be formed in the following way:
\begin{align}
\mathcal{F}\left\{h(r)\right\}
=\mathcal{F}^*+\alpha^*\left( \mathcal{C}\left\{h(r)\right\} + \mathcal{A}\left\{h(r)\right\} \right)
\label{eq_FCA}
\end{align}
where $\mathcal{F}^*$,  $\alpha^*$ are constants with respect to $h(r)$, and where $\mathcal{A}\left\{h(r)\right\}$ is such that the closure relation is equivalent to:
\begin{align}
\frac{\delta \mathcal{A}\left\{h(r)\right\}}{\delta h(r)}=+\rho^2c(r)
\end{align}
If the closure relation has the form $c(r)=\psi\left(h(r);r\right)$, this amounts to requiring:
\begin{align}
\frac{\delta \mathcal{A}\left\{h(r)\right\}}{\delta h(r)}=\rho^2\psi\left(h(r);r\right)
\end{align}
which leads to:
\begin{align}
\mathcal{A}\left\{h(r)\right\}
=\mathcal{A}^*+\rho^2\int_0^1 dt\int d^3r \left\{ h(r)\psi\left(th(r);r\right) \right\}
\end{align}
where $\mathcal{A}^*$ is constant with respect to $h(r)$. In view of Eq.~\eqref{eq_FCA}, we can choose to set $\mathcal{C}^*=\mathcal{A}^*=0$ and keep only $\mathcal{F}^*$ without loss of generality.

We can go a step further than Olivares and McQuarrie towards free-energy functionals by requiring the charging relation to hold at equilibrium. Let us assume a closure relation of the form: $c(r)=-\beta u(r) + \psi'\left(h(r);r\right)$, where $\psi'\left(h(r);r\right)$ does not depend on $u(r)$. Then, if one writes the functional $\mathcal{F}^\xi$ related to a system with interaction potential $\xi u(r)$, one has:
\begin{align}
\mathcal{F}^\xi\left\{h(r)\right\}
=\mathcal{F}^{*\,\xi}&+\alpha^{*\,\xi}
\left(
\int \frac{d^3k}{(2\pi)^3}\left\{\rho h_k - \ln\left(1+\rho h_k\right)\right\}
\right.\nonumber\\
&-\rho^2\beta \int d^3r\left\{h(r)\xi u(r)\right\}
\nonumber\\
&\left.+\rho^2\int d^3r\left\{h(r)\int_0^1 dt \left\{\psi\left(th(r);r\right)\right\} \right\}
\right)
\end{align}
On the other hand, we can recast the charging relation for the renormalized excess free energy, Eq.~\eqref{eq_charging}, as:
\begin{align}
\frac{\partial}{\partial \xi}\left(\frac{A^\xi\left\{h_\text{eq}^\xi(r);\rho,T\right\}}{V}\right)=\frac{\rho^2}{2}\int d^3r
\left\{h_\text{eq}^{\xi}(r)u(r)\right\}
\label{eq_charging_diff}
\end{align}
with the condition that $A_\text{eq}^\xi(\rho,T)/V=0$ if $\xi=0$.
Differentiating the equilibrium value of $\mathcal{F}^\xi$ with respect to $\xi$, we get:
\begin{align}
\frac{\partial \mathcal{F}^\xi\left\{h^\xi_\text{eq}(r)\right\}}{\partial \xi}
=&\left.\frac{\partial \mathcal{F}^\xi\left\{h(r)\right\}}{\partial \xi}\right|_{h^\xi_\text{eq}(r)}\\
=&-\alpha^{*\,\xi}\rho^2\beta \int d^3r\left\{h^\xi_\text{eq}(r) u(r)\right\}
+\frac{\partial \mathcal{F}^{*\,\xi}}{\partial \xi}
\nonumber\\
&+\frac{\partial \alpha^{*\,\xi}}{\partial \xi}
\left(
\int \frac{d^3k}{(2\pi)^3}\left\{\rho h_k - \ln\left(1+\rho h_k\right)\right\}
\right.\nonumber\\
&-\rho^2\beta \int d^3r\left\{h(r)\xi u(r)\right\}
\nonumber\\
&\left.+\rho^2\int d^3r\left\{h(r)\int_0^1 dt \left\{\psi\left(th(r);r\right)\right\} \right\}
\right)
\label{eq_deriv_eq_F}
\end{align}
Thus, the choice $\alpha^{*\,\xi}=-1/(2\beta)$, $\mathcal{F}^{*\,\xi}=0$ is sufficient to fulfill Eq.~\eqref{eq_charging_diff}. This choice leads to the functional of Eq.~\eqref{eq_DH_functional} in the DH case, and to Lado's functional in the HNC case. However, in view of Eq.~\eqref{eq_deriv_eq_F}, this choice is sufficient but not necessary. Finding another choice of $\alpha^{*\,\xi}$,  $\mathcal{F}^{*\,\xi}$ which fulfills Eq.~\eqref{eq_charging_diff} is probably possible but may be cumbersome. However this is mostly a matter of formulation of the problem.

Lets us now consider a slightly different formulation. If one addresses the same problem of finding a functional $\mathcal{F}$ such that the DH equation is equivalent to Eq.~\eqref{eq_mini_F}, but without separating explicitly the OZ relation through the functional $\mathcal{C}$, then one can search for:
\begin{align}
\frac{\delta \mathcal{F}\left\{h(r)\right\}}{\delta h_k}
=\gamma_k^{*}\left(h_k+\beta u_k+\rho\beta h_k u_k\right)
\end{align}
with $\gamma_k^{*}$ independent of $h_k$. This leads to:
\begin{align}
\mathcal{F}\left\{h(r)\right\}
=\mathcal{F}^{*} + \int \frac{d^3k}{(2\pi)^3}\left\{\gamma_k^{*}h_k\left(\frac{h_k}{2}+\beta u_k+\frac{\rho\beta}{2} h_k u_k\right)\right\}
\end{align}
which corresponds to the form that we postulate in Eq.~(11) of \cite{Piron16}. Then, requiring the charging relation to hold at equilibrium leads to our previous result, recalled in Eq.~\eqref{eq_DH_functional_2016}.

\section{Extension to multi-component fluids\label{sec_multicomponent}}
In Ref.~\cite{Olivares76} the authors extend to multi-component fluids the formalism that we use in the discussion of previous section. We can then use their approach in order to extend our free-energy expression of Eq.~\eqref{eq_DH_functional} to multi-component fluids. For that purpose, we define the matrices of functions:
\begin{align}
\bar{\bar{f}}(r)=\left[
\begin{array}{c c c}
  \rho_1 f_{11}(r) & \sqrt{\rho_1\rho_2} f_{12}(r) & ... \\
  \sqrt{\rho_1\rho_2} f_{12}(r) & \rho_2 f_{22}(r) & ... \\
   \vdots  & \vdots & \ddots
\end{array}
\right]
\label{eq_def_matrix}
\end{align}
where $f$ can be the correlation function, the interaction potential or the direct correlation function. The indices label the various species of the multi-component system.
The multi-component OZ relation is:
\begin{align}
c_{ij}(r)=h_{ij}(r)+\sum_{\ell}\rho_\ell\int d^3r'\left\{h_{i\ell}(r')c_{\ell j}(|\vec{r}-\vec{r}'|)\right\}
\end{align}
Using the matrix form of Eq.~\eqref{eq_def_matrix} in the Fourier space, this is equivalent to:
\begin{align}
\bar{\bar{c}}_k=(\mathbb{I}+\bar{\bar{h}}_k)^{-1}\bar{\bar{h}}_k
\end{align}
The functional $\mathcal{C}$ such that:
\begin{align}
\frac{\delta \mathcal{C}\left\{\bar{\bar{h}}(r)\right\}}{\delta \bar{\bar{h}}_k}=-(\mathbb{I}+\bar{\bar{h}}_k)^{-1}\bar{\bar{h}}_k
\end{align}
is given in \cite{Olivares76} as:
\begin{align}
\mathcal{C}\left\{\bar{\bar{h}}(r)\right\}
=\int \frac{d^3k}{(2\pi)^3}\left\{
\ln\left(\det\left(\mathbb{I}+\bar{\bar{h}}_k\right)\right)
-\text{Tr}\left(\bar{\bar{h}}_k\right)
\right\}
\end{align}
where we have set the constant $\mathcal{C}^*$ to zero.

The multicomponent DH closure relation is:
\begin{align}
c_{ij}(r)=-\beta u_{ij}(r)
\end{align}
The functional $\mathcal{A}$ such that:
\begin{align}
\frac{\delta \mathcal{A}\left\{\bar{\bar{h}}(r)\right\}}{\delta \bar{\bar{h}}_{ij}(r)}=-\beta \bar{\bar{u}}_{ij}(r)
\end{align}
can be written as:
\begin{align}
\mathcal{A}\left\{\bar{\bar{h}}(r)\right\}
=-\beta \int d^3r\left\{\sum_{i,j}\bar{\bar{h}}_{ij}(r)\bar{\bar{u}}_{ij}(r)\right\}
\end{align}
where we have set the constant $\mathcal{A}^*$ to zero.
Again, a generating functional for the DH integral equation can be constructed as:
\begin{align}
\mathcal{F}\left\{\bar{\bar{h}}(r)\right\}
=\mathcal{F}^*+ \alpha^*\left(
\mathcal{C}\left\{\bar{\bar{h}}(r)\right\}+\mathcal{A}\left\{\bar{\bar{h}}(r)\right\}
\right)
\end{align}

In the multi-component case, the charging relation reads:
\begin{align}
\frac{\partial}{\partial \xi}\left(\frac{A^\xi\left\{\bar{\bar{h}}_\text{eq}^\xi(r);\rho,T\right\}}{V}\right)=\frac{1}{2}\int d^3r
\left\{\sum_{i,j}\bar{\bar{h}}_{\text{eq},ij}^{\xi}(r)\bar{\bar{u}}_{ij}(r)\right\}
\label{eq_charging_diff_multi}
\end{align}
Like in the one-component case, we can choose $\alpha^{*\,\xi}=-1/(2\beta)$, $\mathcal{F}^{*\,\xi}=0$, which is sufficient to fulfill Eq.~\eqref{eq_charging_diff_multi}.
Finally, we obtain for the DH free-energy functional:
\begin{align}
\frac{A^\text{DH}\left\{\bar{\bar{h}}(r);\rho,T\right\}}{V}
=&\frac{1}{2\beta}\int \frac{d^3k}{(2\pi)^3}\left\{
\text{Tr}\left(\bar{\bar{h}}_k\right)
-\ln\left(\mathbb{I}+\bar{\bar{h}}_k\right)
\right\}\nonumber\\
&+\frac{1}{2}\int d^3r\left\{\sum_{i,j}
\bar{\bar{h}}_{ij}(r)\bar{\bar{u}}_{ij}(r)
\right\}
\label{eq_DH_functional_multicomp}
\end{align}

First, one can readily check that in the one-component case, Eq.~\eqref{eq_DH_functional_multicomp} reduces to Eq.~\eqref{eq_DH_functional}. Moreover, it was shown that the functional of \cite{Blenski17} reduces to that of \cite{Piron16} in the one-component case. Therefore, it is clear that the functional of Eq.~\eqref{eq_DH_functional_multicomp} is distinct from that of \cite{Blenski17}.

On the other hand, in the two-component case, one can compare the equilibrium value of the present functional to that of \cite{Blenski17}. For a two-component fluid, the equilibrium correlation functions are given by:
\begin{align}
\rho_1 h_{\text{eq},11;k}&=\frac{1+\beta \rho_2 u_{22;k}}{D_k}-1\\
\rho_2 h_{\text{eq},22;k}&=\frac{1+\beta \rho_1 u_{11;k}}{D_k}-1\\
\sqrt{\rho_1\rho_2}h_{\text{eq},12;k}&=-\frac{\beta \sqrt{\rho_1\rho_2} u_{12;k}}{D_k}
\end{align}
with:
\begin{align}
D_k=&1+\beta(\rho_1 u_{11;k}+\rho_2 u_{22;k})\nonumber\\
&+\beta^2\rho_1\rho_2(u_{11;k}u_{22;k}-u_{12;k}^2)
\end{align}
Substituting these into Eq.~\eqref{eq_DH_functional_multicomp}, one can check that the equilibrium value of the functional is:
\begin{align}
&\frac{A^\text{DH}\left\{\bar{\bar{h}}_\text{eq}(r);\rho,T\right\}}{V}
\nonumber\\
&=
\frac{1}{2\beta}\int \frac{d^3k}{(2\pi)^3}\left\{
\ln\left(D_k\right)-\rho_1\beta u_{11;k}-\rho_2\beta u_{22;k}
\right\}
\end{align}
which is identical to Eq.~(21) of \cite{Blenski17}. Once again, the identity, in  the sense of thermodynamical functions, of both free-energy functionals taken at equilibrium immediately yields the identity of all thermodynamical functions.

\section{Discussion}
In Eqs.~\eqref{eq_DH_functional} and \eqref{eq_DH_functional_multicomp}, we propose two free-energy functionals that are distinct from the functionals of \cite{Piron16} and \cite{Blenski17}, respectively. The former two however have the same thermodynamical properties as the latter two and, indeed, are identical to them at the DH equilibrium. They only differ out of equilibrium, i.e. when one considers pair distribution functions (or, equivalently, correlation functions) that are not solutions of the DH equation.

This may seem disturbing if one thinks of the density functional theory, in which it is proven that the free-energy functional is a \emph{unique} functional of the density. However, this proof is closely related to the degree of freedom provided by the external potential. In the density functional theory, the uniqueness of the free-energy functional for ``any'' density function relies on two hypotheses:
\begin{itemize}
\item the presence of an external potential in the system, which allows us to associate an ``arbitrary'' density to an external potential
\item the restriction of candidate density functions to the particular class of functions that can emerge as equilibrium densities for systems having suitable external potentials.
\end{itemize}
In the present case, we are dealing with the statistical physics of a homogeneous fluid, which can be addressed using the density functional theory of an inhomogeneous system using the Percus trick \cite{Percus64}. However, this trick implies fixing the external potential to the interaction potential. The density for such a system then becomes the pair distribution function of the considered homogeneous system. 

In this context, the degree of freedom related to the external potential is lost, and it may be that the functional is meaningless out of the equilibrium. If this is the case, having non-unique functionals may be tolerable. Somehow, this would also imply that the pair distribution function cannot be viewed as a physically-relevant internal degree of freedom of the system. Variational approaches with respect to the pair distribution function would then have to be considered as practical mathematical formulations, rather than as physically-motivated ways of deriving the integral equations of fluid models.

\section{Conclusion}
In the present addendum, we supplement the results of our previous publications \cite{Piron16, Blenski17} by proposing alternative Debye-H\"{u}ckel free-energy functionals in both the cases of one-component and multi-component fluids. While resulting in the same thermodynamical relations, these functionals differ when evaluated out of equilibrium. We discuss how these functionals can be obtained following the approach of Lado~\cite{Lado73} as well as that of Olivares and McQuarrie~\cite{Olivares76}. We show that the non-uniqueness issue, which is raised by these results, may be explained through a choice of constants that appear in the method of Olivares and McQuarrie. Finally, we comment briefly on this non-uniqueness issue in the context of the statistical physics of homogeneous fluids.

%

\bibliographystyle{unsrt}
\bibliography{main}

\end{document}